\documentclass{ws-ijmpa}
\usepackage{times}

\newcommand{\E}{E_{0}}
\newcommand{\Eem}{E_{\mathrm{em}}}
\newcommand{\Ecal}{E_{\mathrm{cal}}}
\newcommand{\Xmax}{X_{\mathrm{max}}}

\begin{document}

\markboth{S.~P.~Knurenko et al.}
         {Influence of Primary Cosmic Radiation Mass Composition
          on the Estimation of EAS Energy
         }
\catchline{}{}{}{}{}

\title{INFLUENCE OF PRIMARY COSMIC RADIATION MASS COMPOSITION \\
       ON THE ESTIMATION OF EAS ENERGY
      }

\author{S.~P.~KNURENKO$^*$, A.~A.~IVANOV, V.~A.~KOLOSOV, Z.~E.~PETROV, \\
        I.~YE.~SLEPTSOV and G.~G.~STRUCHKOV
       }
\address{Yu.~G.~Shafer Institute of Cosmophysical Research and Aeronomy
         Lenin Avenue 31, Yakutsk 677891, Russia \\
         ${}^*$s.p.knurenko@ikfia.ysn.ru
        }

\maketitle

\pub{Received (Day Month Year)}{Revised (Day Month Year)}

\begin{abstract}
Energy portion $\Eem / \E$ transferred to EAS electron--photon component at $\E =
10^{15} \pm 10^{19}$~eV was estimated by using of \v{C}erenkov radiation data and charged
particles data obtained at the Yakutsk array. The results are compared with the models for
different energy dissipation into EAS electron--photon component and with calculations
for different primary nuclei. In the energy intervals $10^{15} \pm 10^{16}$~eV and
$10^{18} \pm 10^{19}$~eV $\Eem/\E$ is equal to $(77 \pm 2)$ and $(88 \pm 2)$\%
respectively, that doesn't contradict with mixed composition of primary particles for the
first energy interval and the proton composition for the second one.

\end{abstract}

\section{Introduction}
Energy characteristics of a shower such as the energy lose by particles on ionization of
medium and total energy transferred to electron--photon EAS component are always
considered as important for choice of the interaction mechanism of a cosmic ray primary
particle with nuclei of atoms of the air. And these characteristics can be used for
estimation of primary particle energy without a definite model. The energy can be
estimated only in two cases (without using of any notions of the interaction model): when
the total flux of \v{C}erenkov light at the sea level has been measured; the longitudinal
EAS development, a more exactly, the total number of particles in the maximum of shower
development has been measured.  At present time, only at two EAS arrays such measurements
are carried out: at the Yakutsk array (Russia) where the \v{C}erenkov EAS radiation is
measured and at the HiRes array (USA) where measurements of the ionization nitrogen
luminosity are provided.

In the present work, a problem of determination of the energy portion transferred to
electromagnetic EAS cascade is studied. The wide energy interval from $10^{15}$ to
$10^{19}$~eV is considered. In the method of determination of $\Eem$ the real conditions
of the atmosphere, the longitudinal development of a shower $(\Xmax)$ and the mass
composition of primary particles are taken into account.

\section{Energy Transferred to the Electromagnetic EAS Component}

\begin{figure}[htb]
\centering
\includegraphics[width=0.9\linewidth]%
 {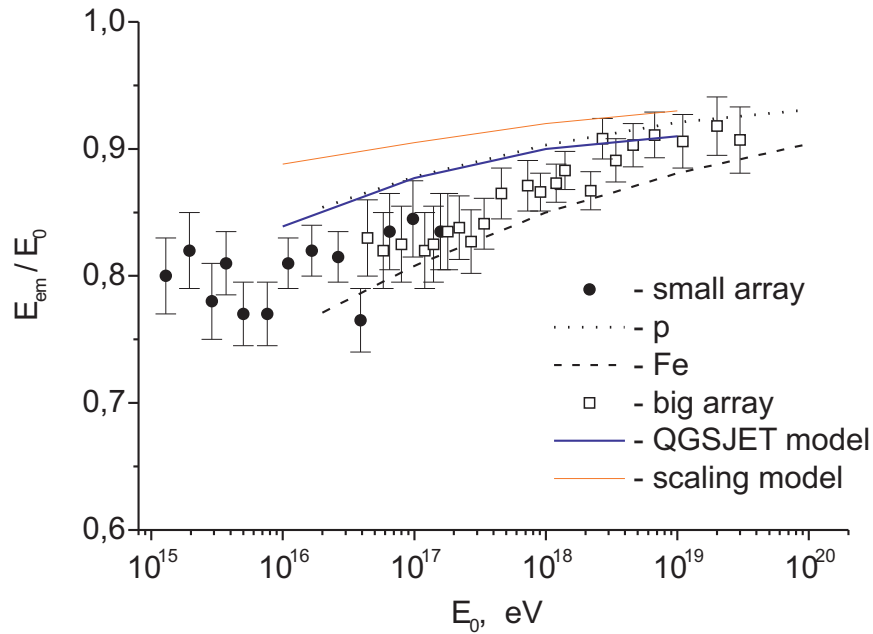}
\vspace*{8pt}
\caption{A portion of the energy transferred to the electromagnetic
         EAS component by \v{C}erenkov light data at the Yakutsk array.}
\label{fig1}
\end{figure}

Figure~\ref{fig1} presents the Yakutsk EAS array experimental data together with
calculations by the model with decelerated and moderate dissipation of energy into
electromagnetic EAS component: quasiscaling models (solid line) and QGSJET (dashed
line)\cite{bib1}.
From Fig.~\ref{fig1} one can see both the agreement of experimental data with calculations
by QGSJET model (proton) in the region $\E \ge 3 \times 10^{18}$~eV, and disagreement at
$\E \le 3 \times 10^{18}$~eV. The scaling model gives a noticeably greater value of $\Eem
/ \E$ in relation to the experimental data that is doubtlessly also connected with the
break of scaling function in the region of ultra--high energies.

The experimental data in Fig.~\ref{fig1} is well approximated by the expression of a form:
\begin{equation}
\Eem / \E = (0.964 \pm 0.011) - (0.079 \pm 0.005) \times \E^{-(0.147 \pm 0.008)}\mathrm{,}
\label{eq1}
\end{equation}
where $\E$ is taken in units of eV.

The relation (\ref{eq1}) is primarily important for the comparison of estimations of $\E$
obtained at the Yakutsk and Fly's Eye arrays. The new estimation of $\Ecal / \E$ parameter
for the Fly's Eye array is given in Ref.~\refcite{bib2}:
\begin{equation}
\Ecal / \E = (0.959 \pm 0.003) - (0.082 \pm 0.003) \times \E^{-(0,147 \pm 0,006)}
\label{eq2}
\end{equation}

From (\ref{eq1}) and (\ref{eq2}) a good coincidence of both results, calculation and
empirical estimation is well seen. The calculations in Ref.~\refcite{bib1} have been carried out
by using of QGSJET model in the case of primary proton and iron nucleus (see Table~1).
\begin{table}[htb]
\tbl{Yakutsk data and calculations for p$/$Fe}
{\begin{tabular}{@{}ccccc@{}}
\toprule
      &           &    experiment     &  QGSJET model   & SIBYLL model    \\
$n/n$ & $\E$, eV  &  $\Eem / \E$      & $\Eem / \E$     & $\Eem / \E$     \\
\colrule
1     & $10^{17}$ & $0.824 \pm 0.015$ & $0.876 / 0.806$ & --              \\ 
2     & $10^{18}$ & $0.872 \pm 0.008$ & $0.901 / 0.850$ & $0.915 / 0.865$ \\
3     & $10^{19}$ & $0.911 \pm 0.010$ & $0.914 / 0.879$ & $0.934 / 0.897$ \\
\botrule
\end{tabular}}
\end{table}
In the case of the primary proton a good agreement between both calculations is observed.
The comparison of experimental data with calculations for the proton and iron nucleus
indicates the fact that the mass compositions of cosmic radiation particles in the
energy region of $10^{17}-10^{19}$~eV and above $3\times10^{18}$~eV must differ.
At $\E\ge3\times10^{18}$~eV the mass composition is most likely close to the proton one.
Therefore, when estimating the EAS energy at the Fly's Eye array it is more reasonably to
use the formula (\ref{eq2}). At the Yakutsk array the formula (\ref{eq1}) obtained
empirically is used. A systematic difference between (\ref{eq1}) and (\ref{eq2}) does not
exceed 10\% that speaks for a coincidence of estimations of $\E$ at the Yakutsk and Fly's
Eye arrays.

\section*{Acknowledgements}

This work has been financially supported by RFBR, grant \#02--02--16380, grant
\#03--02--17160 and grant INTAS \#03--51--5112.


\begin{thebibliography}{0}
\bibitem{bib1}
S.~P.~Knurenko {\it et al}., in {\it Proc.\ 26th Int.\ Cosmic Ray Conf.,
Salt Lake City} {\bf 1}, 372 (1999).
\bibitem{bib2}
C.~Song, Z.~Cao, B.~R.~Dawson {\it et al}., {\it Astropart.\ Phys.\/}
{\bf 14}, 7 (2000).
\end{thebibliography}
\end{document}